\documentstyle[epsfig]{aprim}

\newif\ifAMStwofonts

\ifoldfss
  \ifCUPmtlplainloaded \else
    \NewTextAlphabet{textbfit} {cmbxti10} {}
    \NewTextAlphabet{textbfss} {cmssbx10} {}
    \NewMathAlphabet{mathbfit} {cmbxti10} {} 
    \NewMathAlphabet{mathbfss} {cmssbx10} {} 
  \fi
  \ifAMStwofonts
    \ifCUPmtlplainloaded \else
      \NewSymbolFont{upmath} {eurm10}
      \NewSymbolFont{AMSa} {msam10}
      \NewMathSymbol{\upi}     {0}{upmath}{19}
      \NewMathSymbol{\umu}     {0}{upmath}{16}
      \NewMathSymbol{\upartial}{0}{upmath}{40}
      \NewMathSymbol{\leqslant}{3}{AMSa}{36}
      \NewMathSymbol{\geqslant}{3}{AMSa}{3E}

    \fi
  \fi
\fi 

\ifnfssone
  \newmathalphabet{\mathit}
  \addtoversion{normal}{\mathit}{cmr}{m}{it}
  \addtoversion{bold}{\mathit}{cmr}{bx}{it}
  \newmathalphabet{\mathbfit} 
  \addtoversion{normal}{\mathbfit}{cmr}{bx}{it}
  \addtoversion{bold}{\mathbfit}{cmr}{bx}{it}
  \newmathalphabet{\mathbfss} 
  \addtoversion{normal}{\mathbfss}{cmss}{bx}{n}
  \addtoversion{bold}{\mathbfss}{cmss}{bx}{n}
  \ifAMStwofonts
    \ifCUPmtlplainloaded \else
      \UseAMStwoboldmath
      \makeatletter
      \new@mathgroup\upmath@group
      \define@mathgroup\mv@normal\upmath@group{eur}{m}{n}
      \define@mathgroup\mv@bold\upmath@group{eur}{b}{n}
      \edef\UPM{\hexnumber\upmath@group}
      \new@mathgroup\amsa@group
      \define@mathgroup\mv@normal\amsa@group{msa}{m}{n}
      \define@mathgroup\mv@bold\amsa@group{msa}{m}{n}
      \edef\AMSa{\hexnumber\amsa@group}
      \makeatother
      \mathchardef\upi="0\UPM19
      \mathchardef\umu="0\UPM16
      \mathchardef\upartial="0\UPM40
      \mathchardef\leqslant="3\AMSa36
      \mathchardef\geqslant="3\AMSa3E
    \fi
  \fi
\fi 

\ifnfsstwo
  \DeclareMathAlphabet{\mathbfit}{OT1}{cmr}{bx}{it}
  \SetMathAlphabet\mathbfit{bold}{OT1}{cmr}{bx}{it}
  \DeclareMathAlphabet{\mathbfss}{OT1}{cmss}{bx}{n}
  \SetMathAlphabet\mathbfss{bold}{OT1}{cmss}{bx}{n}
  \ifAMStwofonts
    \ifCUPmtlplainloaded \else
      \DeclareSymbolFont{UPM}{U}{eur}{m}{n}
      \SetSymbolFont{UPM}{bold}{U}{eur}{b}{n}
      \DeclareSymbolFont{AMSa}{U}{msa}{m}{n}
      \DeclareMathSymbol{\upi}{0}{UPM}{"19}
      \DeclareMathSymbol{\umu}{0}{UPM}{"16}
      \DeclareMathSymbol{\upartial}{0}{UPM}{"40}
      \DeclareMathSymbol{\leqslant}{3}{AMSa}{"36}
      \DeclareMathSymbol{\geqslant}{3}{AMSa}{"3E}
    \fi
  \fi
\fi 

\ifCUPmtlplainloaded \else
  \ifAMStwofonts \else 
    \def\upi{\pi}
    \def\umu{\mu}
    \def\upartial{\partial}
  \fi
\fi

\title[Planetary Systems]
 {The Roles of Discs for Planetary Systems}

\author[Yeh and Jiang]
       {Li-Chin Yeh$^1$ and Ing-Guey Jiang$^2$\\
  $^1$Department of Applied Mathematics, 
      National Hsinchu University of Education, Hsin-Chu, Taiwan\\
  $^2$Institute of Astronomy, National Central University, Chung-Li, Taiwan}
\date{}

\pagerange{\pageref{firstpage}--\pageref{lastpage}}
\pubyear{2005}

\begin{document}

\maketitle

\label{firstpage}

\begin{abstract}

It is known that the discs are detected for some of the extra-solar
planetary systems. It is also likely that there was a disc mixing with planets
and small bodies while our Solar System was forming. From our recent results,
we conclude that the discs play two roles: the gravity makes planetary systems
more chaotic and the drag makes planetary systems more resonant.

\end{abstract}


\section{Introduction}

More than 160 extra-solar planetary systems have been discovered
so far and the number of detected exoplanets is  
likely to increase rapidly. This development has opened a new window for
astronomers to investigate the planetary formation and evolution in  
detail.
There are many numerical studies on the dynamical evolution of these detected 
systems
(please see Laughlin \& Adams 1999, Rivera \& Lissauer 2000, Jiang \& Ip 2001,
Ji et al. 2002, Benest 2003). 

Among these systems, the existence of discs seems
to be a general phenomenon. For instance,
there are asteroid belt and Kuiper belt for the Solar System, 
the discs of dust for extra-solar planetary systems and also circumbinary
rings for binary systems.  
The disc or belt-like structure should influence the dynamical evolution of 
these systems.
As mentioned in
Jiang \& Ip (2001), the origin of orbital elements of  
the planetary system of upsilon Andromedae
might relate to the disc interaction.
Moreover, Yeh \& Jiang (2001) studied the orbital 
migration of scattered planets. They completely classify the 
parameter space and solutions
and conclude that the eccentricity always increases if the planet, which 
moves on circular orbit initially, is scattered to migrate outward.
Thus, the orbital circularization must be important for scattered planets
if they are now moving on nearly circular orbits.

To concentrate on the dynamical role of discs for the planetary systems,
Jiang \& Yeh (2003, 2004a) did some analysis
on the orbital evolution for systems with 
planet-disc interaction.


\section{The Chaotic Orbits}
Jiang \& Yeh (2004b) 
studied the chaotic orbits for the disc-star-planet systems
through the calculations of Lyapounov Exponent. 
Different initial conditions are used in the 
numerical surveys to explore the possible chaotic and regular orbits. 
They found that, in general, 
discs with different masses might change the sizes and locations 
of chaotic region. 
Some sample orbits are further studied
and the plots of Poincar\'e
surface of section are consistent with the results of Lyapounov Exponent
Indicator.
On the other hand, they pointed out  
that the influence of the disc 
can change the locations of equilibrium points and also the orbital
behaviors. 
This point is particularly interesting. Without the influence from the 
disc, Wisdom (1983) showed that one of the Kirkwood gaps
might be explained by the chaotic boundary. This is probably 
due to that the chaotic orbits are less likely to become resonant orbits.
Indeed, the chaotic orbits in the disc-star-planet system 
shall play essential
roles during the formation of Kirkwood gaps. 
For example, 
the disc might make some test particles to transfer from chaotic orbits
to regular orbits
or from regular orbits to chaotic orbits
during the evolution.

The discs are often observed with various masses in both 
the young stellar systems and extra-solar 
planetary systems.
It is believed that the proto-stellar discs are formed during the early stage
of star formation and the planets might form on the discs through 
the core-accretion or disc instability mechanism. The planetary 
formation timescale is not clear but the gaseous disc will be somehow 
depleted and 
the dust particles will gradually form. 
The above picture is plausible but unfortunately the timescales are not
completely understood. For a system with a star, planets, discs and asteroids, 
it is not clear what the order of different component's 
formation is and what the duration of each component's formation process
would be.

Nevertheless, one thing we can confirm is that the discs are there and
can have different masses at different stages. From our results, it is clear
that the dynamical properties of orbits, i.e. chaotic or regular,  
can be determined quickly 
for different disc masses or models. Because those particles moving
on chaotic orbits might have smaller probability to become the resonant objects
in the system, the Kirkwood gaps, which are associated with 
the location of mean motion resonance, in the Solar System might be formed 
by the dynamical effect we have discussed here.
The detail processes would depend on the evolutionary 
timescale of the star, planet and also the disc.
To conclude, it is important that the effect of discs on the chaotic orbits
might influence the formation of the structures of planetary systems such as
the Kirkwood gaps in the Solar System.

\section{The Resonance}


The mean motion resonant relations between different celestial bodies
are confirmed to be common phenomena 
for both Solar System and extra-solar planetary systems.  
For example, the gaps in the distribution of asteroids were connected to 
mean motion resonances by Kirkwood (1867) and known as Kirkwood gaps.
Greenberg \& Scholl (1979) classified the proposed explanations 
 into four groups:
the statistical hypothesis, the collisional hypothesis,
the cosmogonic hypothesis, and the gravitational hypothesis.
Dermott \& Murray (1983) showed that there is a good correspondence between 
the width of the Kirkwood gaps and the width of the libration zone when the
distribution of asteroid orbits is plotted in the $a-e$ plane. They concluded
that the gravitational hypothesis was likely to be correct.
This is completely consistent with the results on the chaotic orbits as we
mentioned in the last section (Jiang \& Yeh 2004b).


In addition to that, the small bodies at the outer Solar System, i.e. 
the Kuiper Belt Objects (KBOs), 
also show strong mean motion resonant relations with
the Neptune's orbit. It is surprising that about one-third of the 
KBOs are engaged into 3:2 resonance and are therefore called ``Plutinoes''.
The sweeping mechanism proposed in Malhotra (1995) that the outward migration
of Neptune might make the Plutinoes catched into 3:2 resonance. This 
mechanism could work under the condition that the
Neptune's orbit keep changing and thus the Neptune's potential felt by the 
Plutinoes is time-dependent.  This relative orbital movement might become
negligibly small when the Neptune and KBOs are pushed out together.
This scenario of pushing-out 
from the inner Solar System proposed by Levison \& Morbidelli (2003)
was to solve the KBOs' formation issue since the density was higher
in the inner Solar System and easier to form KBOs (and also Neptune).
Moreover, Yeh \& Jiang (2001) showed that the Neptune would be in an 
eccentric orbit if it was scattered outward. It will need a massive 
belt to circularized its orbit.
Therefore, it is not clear that how the Plutinoes was captured into 3:2 
resonance under the situation that the formation history of KBOs is unclear.
Jiang \& Yeh (2004c) proposed that the drag-induced resonant capture
could play a role for this important process. 
(Please also see a brief review in Jiang \& Yeh 2005.)

On the other hand, some extra-solar multiple planetary systems also 
show the resonances. For example, Ji et al. (2003) confirmed that 
the 55 Cancri planetary system is indeed in the 3:1 mean motion resonance
by both the numerical simulations and secular theory. The 
GJ 876 and HD 82943 planetary systems are probably in 2:1 resonance
as studied by Laughlin \& Chambers (2001), Kinoshita \& Nakai (2001),
Ji et al. (2002) and Gozdziewski \& Maciejewski (2001).

\section{Concluding Remarks}

The conventional models used in celestial mechanics consider 
the dynamical interactions between particles only. 
We take a first step to include the dynamical effect from the disc in
the conventional celestial mechanics.
In general, the disc shall vary with time. 
Nevertheless, in these studies, we only
consider the orbital behaviors during a timescale shorter than the
lifetime of the disc, the models shall give good implications
for the first trend of orbital evolution under the
disc potential. We find that when we consider the gravity only,
there are more chaotic orbits when the disc is included.
We also consider the effect of an artificial drag and find that 
the drag could bring particles into resonances. 
Our results' general conclusion is consistent with the ones 
in Armitage et al. (2002) and Trilling et al. (2002),
though they have different focuses, base on different approaches,
and mainly study  the gas-dynamical effects.

\section*{Acknowledgment}
We are grateful to the National Center for High-performance Computing
for computer time and facilities.

\label{lastpage}

\clearpage

\end{document}